\documentclass{article}
\usepackage{spconf,amsmath,graphicx,hyperref}
\usepackage{multicol,multirow, tabularx}

\title{Articulation-Informed ASR: Integrating Articulatory Features into ASR via Auxiliary Speech Inversion and Cross-Attention Fusion}
\name{Ahmed Adel Attia\textsuperscript{1},  Jing Liu\textsuperscript{2}, Carol Espy Wilson\textsuperscript{1}}
\address{University of Maryland\\
\textsuperscript{1}Department of Electrical and Computer Engineering, \textsuperscript{2}College of Education\\
aadel@umd.edu, jliu28@umd.edu, espy@umd.edu}%
\begin{document}
%
\maketitle

\begin{abstract}
Prior works have investigated the use of articulatory features as complementary representations for automatic speech recognition (ASR), but their use was largely confined to shallow acoustic models. In this work, we revisit articulatory information in the era of deep learning and propose a framework that leverages articulatory representations both as an auxiliary task and as a pseudo-input to the recognition model. Specifically, we employ speech inversion as an auxiliary prediction task, and the predicted articulatory features are injected into the model as a query stream in a cross-attention module with acoustic embeddings as keys and values. Experiments on LibriSpeech demonstrate that our approach yields consistent improvements over strong transformer-based baselines, particularly under low-resource conditions. These findings suggest that articulatory features, once sidelined in ASR research, can provide meaningful benefits when reintroduced with modern architectures.
\end{abstract}
\vspace{-10pt}
\section{Introduction}
\vspace{-10pt}
\label{sec:intro}
Vocal Tract Variables (TVs) are trajectories that show the motion and position of the articulators used to produce speech. These signals have been shown to be a very effective representation of speech in a number of fields, including speech pathology \cite{benway2025perceptual, benway2025subtyping}, mental health diagnosis \cite{10.1145/3462244.3479967, siriwardena2021inverted, 11011243, premananth2025multimodal}, speech synthesis \cite{ speech_synthesis_1, speech_synthesis_2}, and relevantly, Automatic Speech Recognition (ASR) \cite{mitra2010articulatory, mitra2017hybrid}. 

In speech recognition, previous works \cite{mitra2010articulatory, mitra2017hybrid} have shown TVs to be strong and robust input signals, particularly under noise. However, these works have been confined to shallow feed-forward networks or Convolutional Neural Networks (CNNs). Additionally, these works only worked with synthetic rather than natural speech. However, they demonstrated that TVs, being less redundant and noisy than acoustics, are effective for ASR. The motion of different parts of the vocal tract (lips, tongue tip, tongue body, etc) is a much more concise representation of speech than the acoustic signal. 

Despite that promise, TVs have seen little use in modern transformer-based ASR models such as Whisper \cite{radford2023robust} and Wav2vec2.0 \cite{baevski2020wav2vec}. These models require large amounts of training data. On the other hand, the process of collecting articulatory data involves the use of technologies like X-Ray Microbeam (XRMB) \cite{Westbury1994a}, or Electromagnetic articulography (EMA) \cite{Schonle1987}. These recording sessions are expensive, and long sessions can be uncomfortable and unsafe for the subjects. As a result, articulatory datasets are only a few hours long, which is not enough to train transformer ASR models. 

Acoustic-to-Articulatory Speech Inversion (SI) is the process of extracting the TVs from the acoustic signal \cite{parikh2022acoustic}. Using SI models, one can theoretically obtain enough TV data to train large transformer models. However, when attempting to train a pure articulatory ASR model, one is faced with another obstacle. Training a large transformer ASR model, like Whisper or Wav2vec2.0, from scratch is expensive and time-consuming, which has unfortunately become out of reach for many academic institutions. Many researchers rely on fine-tuning pre-trained off-the-shelf models to side-step the massive resources required to train these models from scratch. Using pre-trained models, however, prevents any changes in the input side of the model, because doing so shifts the distribution of the activations for later layers. 

Given the scarcity of articulatory data and the impracticality of retraining large ASR models from scratch, we propose a Multi-Task Learning (MTL) approach that incorporates articulatory information into pre-trained ASR models. Siriwardena et. al \cite{siriwardena2022acoustic} proposed an MTL approach to improve the performance of SI systems by adding an auxiliary ASR task. Our method, on the other hand, based on Wav2vec2.0, uses SI as an auxiliary task to encourage articulation-aware representations, while injecting predicted TVs back into the model through a cross-attention module prior to CTC decoding. Experiments on varying amounts of LibriSpeech \cite{7178964} demonstrate consistent improvements, particularly under low-resource and noisy conditions.

\section{Methodology}
\begin{figure}[h!]
    \centering
    \includegraphics[width=\linewidth]{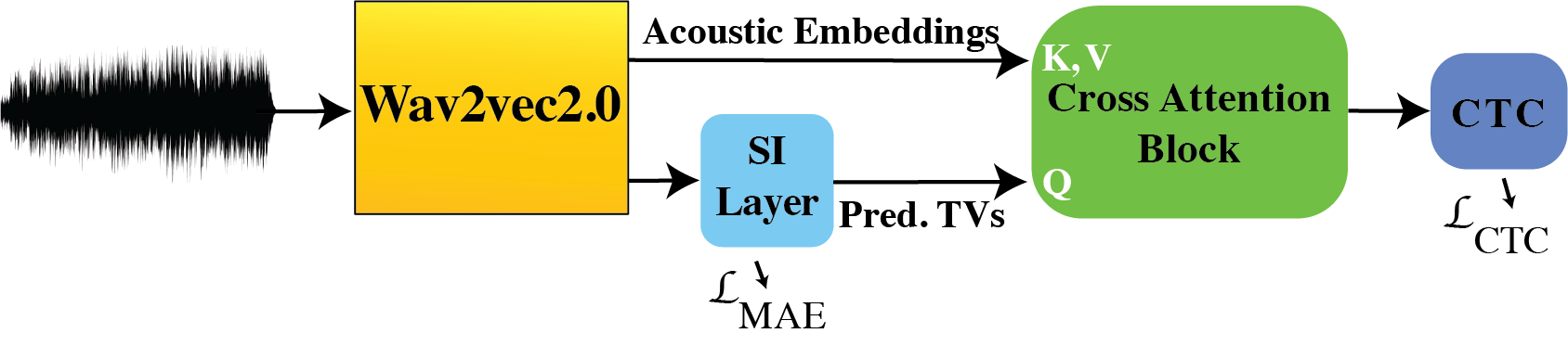}
    \vspace{-20pt}
    \caption{The proposed Articulation-Informed ASR framework.}
    \vspace{-20pt}
    
    \label{fig:model}
\end{figure}

We propose Articulation-Informed ASR, a framework that integrates articulatory information into speech recognition models through a dual strategy: (1) an auxiliary SI task that encourages articulation-aware acoustic speech representations, and (2) a cross-attention fusion block that injects predicted articulatory trajectories as an additional input stream before CTC decoding. The entire model is trained through a MTL paradigm. While we base our design on Wav2vec2.0, our methods here can be applied to ASR models, like Whisper. We leave that implementation to future work. An overview of our model is shown in Figure \ref{fig:model}.
\vspace{-10pt}
\subsection{Auxiliary SI Task} 
\vspace{-10pt}

 Given an acoustic signal, the Wav2vec2.0 transformer extracts contextual speech embeddings. In vanilla Wav2vec2.0, these embeddings are directly classified into graphemes. We follow a similar approach in our model, where we add a single SI feed-forward layer that maps the embeddings into TVs. This task is learned using a simple Mean Absolute Error (MAE) loss between the predicted and target TVs.  The MAE loss trains the SI layer, and also contributes to updating the Wav2vec2.0 backbone. This allows for positive transfer between the SI and ASR tasks. 
\vspace{-10pt}

\subsection{Cross-Attention Fusion for Acoustic-Articulatory CTC}
\vspace{-10pt}
 
To further tie the articulatory information with the ASR task, we add a cross-attention block, otherwise identical to the self-attention blocks in Wav2vec2.0 transformers. The difference is that the keys and values are derived from the Wav2vec2.0 acoustic embeddings, and the queries are derived from the predicted TVs. This weighs the acoustic embeddings based on the agreement between the acoustic embeddings and the TVs. This late fusion of the TVs with the deep acoustic embeddings results in a mixture of top-down and bottom-up representations for the ASR task. The resulting representations are directly fed into the CTC layer, similar to vanilla Wav2vec2.0. This task is learned using the CTC loss, and the gradients backpropagate into the CTC and SI layers, cross-attention block and the Wav2vec2.0 backbone.

\subsection{Loss Functions}
Multi-task learning often requires manually tuning individual task losses and their contributions to the total loss:
\begin{equation}
\mathcal{L}_{t} = \alpha_{\text{CTC}} \, \mathcal{L}_{\text{CTC}} + \alpha_{\text{MAE}} \, \mathcal{L}_{\text{MAE}}.
\end{equation}

The main objective of this tuning process is to ensure training stability and adequate learning for each task. However, in early experiments we found this approach rigid, as weights that work in early stages may not be optimal later.

Uncertainty-based weighting (UBW) provides a more adaptive method of combining regression and classification losses \cite{kendall2018multi}. Instead of hand-crafted weights, UBW introduces two learnable parameters that adjust dynamically based on the progression of each loss. In our experiments, this method outperformed manual weighting, though it proved somewhat sensitive to initialization.

For classification (CTC) and regression (MAE), the combined objective is:
\begin{equation}
\mathcal{L}_{t} = \frac{1}{\sigma_{\text{CTC}}^{2}} \, \mathcal{L}_{\text{CTC}} 
+ \frac{1}{2\sigma_{\text{MAE}}^{2}} \, \mathcal{L}_{\text{MAE}} 
+ \log \sigma_{\text{CTC}} + \log \sigma_{\text{MAE}},
\end{equation}
where $\sigma_{\text{CTC}}$ and $\sigma_{\text{MAE}}$ are learnable parameters corresponding to task uncertainty.

\section{Datasets and Addressing Articulatory Data Scarcity}
Training our multi-task model requires a speech dataset with both articulatory labels and orthographic transcriptions. However, articulatory datasets are typically limited in size and often lack accurate orthographic transcriptions, while ASR datasets do not contain articulatory labels. To combine both tasks, one must either use a pre-trained ASR system to label articulatory datasets with transcriptions or a pre-trained SI system to generate articulatory features for ASR datasets. We adopt the latter approach, as our primary focus is the ASR task, and it is preferable to retain high-quality human transcriptions. Moreover, articulatory datasets span only a few hours, compared to ASR datasets that span hundreds or thousands of hours. Our approach is also motivated by prior work \cite{premananth2025multimodal}, which has shown that machine-predicted articulatory labels can still provide effective supervision in downstream tasks. 

We conduct experiments on LibriSpeech, a high-quality ASR corpus of read speech. A pre-trained SI system \cite{tabatabaee2025enhancing} is employed to predict TVs for the LibriSpeech corpus. The system predicts a set of TVs, including Lip Aperture (LA), Lip Protrusion (LP), Tongue Body Constriction Location (TBCL), Tongue Body Constriction Degree (TBCD), Tongue Tip Constriction Location (TTCL), Tongue Tip Constriction Degree (TTCD), and Velopharyngeal Port (VP). In addition, it extracts two source features: Periodicity and Aperiodicity. In total, we use nine TVs. 

The predicted TVs are sampled at 100 Hz. Wav2Vec2.0 representations are sampled at 50 Hz, matching the 20 ms stride of Wav2Vec2.0. To simplify alignment in the cross-attention module, we downsample the TVs to 50 Hz, ensuring temporal consistency
\section{Experiments}
\label{sec:experiments}

\subsection{Experimental Setup}
We evaluate our proposed method on the LibriSpeech corpus using both Wav2Vec2.0 \textit{base} and \textit{large} models to assess scalability across encoder sizes. All models are fine-tuned under a multi-task setting, where SI is used as an auxiliary task and articulatory features are fused via cross-attention prior to CTC decoding. We also train otherwise identical single-task vanilla Wav2vec2.0 models with the same data and hyperparameters.

To investigate data efficiency, we consider subsets of \{10 minutes, 1 hour, 10 hours, 100 hours\} of labeled training data from the \texttt{train-clean-100} partition of LibriSpeech, following common practice in low-resource ASR evaluations. This setup allows us to examine the contribution of articulatory features under varying amounts of supervision.

The 10-minute and 1-hour models were trained for 13k steps with a learning rate of $5\times10^{-5}$, the 10-hour models were trained for 20k steps with the same rate, and the 100-hour models were trained for 80k steps with a learning rate of $3\times10^{-5}$. All models were trained on A100 or H100 GPUs. 

\vspace{-10pt}
\subsection{Evaluation}
We evaluate our models on \texttt{test-clean} and \texttt{test-other} sets using Word Error Rate (WER). While the SI task is typically evaluated using the Pearson Product-Moment Correlation (PPMC), we focus here on ASR performance. We note that all models achieve high correlation with the machine-generated labels, although none outperform the dedicated SI model \cite{tabatabaee2025enhancing} on real articulatory data.

\vspace{-10pt}
\subsection{Language Model Beam-Search Decoding}
We report results with and without $n$-gram Language Model (LM) beam-search decoding, following \cite{baevski2020wav2vec}. Results without LM integration reflect the raw model performance, while results with LM reflect performance in more practical settings. 
\section{Results}

\begin{table*}[t]
\centering
\caption{WER (\%) on LibriSpeech with Wav2Vec2.0 \textit{Base} and \textit{Large}. Baseline = CTC only; Proposed = SI+ASR with cross-attention fusion. Values shown as No LM / LM.} \vspace{10pt}
\label{tab:base_large_results}
\begin{tabular}{c|c|c|c|c}
\hline
\multirow{2}{*}{Hours} & \multicolumn{2}{c|}{Test-Clean} & \multicolumn{2}{c}{Test-Other} \\ \cline{2-5}
 & Baseline & Proposed & Baseline & Proposed \\
\hline
\multicolumn{5}{c}{\textbf{Base}} \\
\hline
10m  & 57.86 / 50.85 & \textbf{45.83 / 35.83} & 69.52 / 63.74 & \textbf{56.68 / 46.28} \\
1h   & 23.66 / 19.46 & \textbf{20.42 / 17.70} & 35.98 / 30.53 & \textbf{31.37 / 27.92} \\
10h  &  9.75 /  7.59 & \textbf{ 9.05 /  5.57 }& 20.20 / 16.53 & \textbf{18.75 / 15.10} \\
100h &  5.53 /  4.46 &  \textbf{5.43 /  4.32 }& 14.28 / 11.85 & \textbf{12.80 / 10.38} \\
\hline
\multicolumn{5}{c}{\textbf{Large}} \\
\hline
10m  & 39.21 / 35.01 & \textbf{37.94 / 31.94} & 51.03 / 46.73 & \textbf{45.66 / 39.40} \\
1h   & 20.32 / 17.56 & \textbf{16.07 / 13.62} & 30.08 / 26.51 & \textbf{23.69 / 20.43} \\
10h  &  8.62 /  7.46 &  \textbf{6.59 /  6.90} & 16.26 / 12.79 & \textbf{13.58 / 11.96} \\
100h &  4.00 /  \textbf{3.28} &  \textbf{3.89} /  3.41 &  9.56 /  8.22 &  \textbf{9.17 /  8.08} \\
\hline
\end{tabular}
\vspace{-10pt}
\end{table*}

\vspace{-10pt}
\subsection{Analysis}
Overall, the results demonstrate that incorporating articulatory supervision through our multi-task model improves ASR performance across both the \textit{base} and \textit{large} Wav2Vec2.0 configurations. In every training regime, the proposed method either matches or outperforms the baseline, underscoring the robustness of the approach.

For the \textit{base} model, the general trend is that the largest relative gains appear in the data-constrained settings. When training with only 10 minutes of labeled speech, the proposed model produces substantial improvements, with relative WER reductions \textbf{20.79\%} and \textbf{29.52\%} for the \texttt{test-clean} testset with and without LM respectively, and \textbf{18.47\%} and \textbf{27.39\%} for the \texttt{test-other} testset with and without LM respectively. This suggests that the articulatory information is particularly valuable when the acoustic encoder alone cannot generalize well from such limited data and that the added supervision increases the utility of small datasets. 

At 1 hour and 10 hours of training data, we continue to observe consistent improvements across both \texttt{test-clean} and \texttt{test-other}, showing that articulatory features remain beneficial in the low-resource regime. Even at 100 hours, where the baseline is already strong, the multi-task model still provides measurable gains. Specifically, on the \texttt{test-clean} test-set we see a relative improvement of \textbf{1.81\%} and \textbf{3.09\%} with and without an LM. For \texttt{test-other}, we observe a \textbf{10.33\%} relative improvement without an LM and a \textbf{12.41\%} relative improvement with LM decoding. These results indicate that the proposed approach is not only effective in extremely low-resource settings, but also continues to provide complementary information at larger scales.

The story is somewhat different for the \textit{large} model. Here, improvements are observed across most training conditions, but the trend is less uniform. At the lowest data regime of 10 minutes, the relative improvements are modest compared to the \textit{base} configuration. This outcome is expected, since the larger encoder capacity allows the model to extract stronger acoustic representations even with limited data, reducing its dependence on auxiliary supervision. However, as we increase the amount of labeled training data, the articulatory supervision begins to play a more important role. At 1 hour and 10 hours, the \textit{large} configuration with the multi-task model shows clear and consistent gains across both test sets, highlighting that articulatory features can reinforce and stabilize training once sufficient supervision is available. This pattern suggests that the utility of articulatory supervision interacts with model capacity: smaller encoders benefit most when data is scarce, while larger encoders leverage the supervision more effectively in mid-resource regimes.

One outlier evident in the results is that the \textit{large} model with LM decoding shows a slight degradation on \texttt{test-clean}, from 3.28\% WER with the baseline to 3.41\% with the proposed model. However, since the proposed model still improves without LM decoding and also on \texttt{test-other}, we attribute this anomaly to suboptimal LM decoding rather than a fundamental limitation of our method. It is also worth noting that our LM decoding results for the baseline model do not fully match those reported in the seminal work \cite{baevski2020wav2vec}, which may further explain this discrepancy.

Taken together, these findings suggest that the proposed method provides robust improvements in both small and large model configurations, with the nature of the improvements depending on the amount of available labeled data. In the \textit{base} model, articulatory supervision provides strong gains in extremely low-resource conditions, demonstrating its ability to supplement weak acoustic representations. In the \textit{large} model, improvements are less dramatic at the lowest data scales, but grow more consistent as more data becomes available. The overall takeaway is that articulatory supervision is particularly effective in low- and mid-resource regimes, particularly with smaller models, while still offering measurable benefits even when training with larger amounts of data.

An interesting pattern emerges when comparing the proposed MTL \textit{base} model with the \textit{large} baseline, particularly in the medium-resource settings (1 and 10 hours) and, to a lesser extent, in the low-resource regime. The proposed \textit{base} model, closely matches the performance of the much larger \textit{large} baseline model, despite having roughly one third of the parameters. In other words, incorporating articulatory information as both an auxiliary task and an additional input stream effectively narrows the capacity gap between \textit{base} and \textit{large}, yielding performance comparable to a model three times its size with only minimal added complexity.
\vspace{-10pt}
\subsection{Qualitative Example}
To illustrate the effect of articulatory supervision beyond quantitative WER scores, we present a read speech example. \\\\
\textit{\textbf{Ground Truth}:``i can say this is thomas gibbs gee my one and only child and when he finished high school we had always planned to send him to princeton but his father had been called back into the service as a reserve officer and was stationed in washington''}  

At 10 hours of training data, the \textit{base} model without LM decoding produced the following output:  \\\\
\textbf{Baseline:}  
\textit{``i can say this is toms gibsgi mynonly choildand when he finishd hyscol we had always pland to send tim to prinsto but his father had been called back into the servecs as a reserve offiseor and was statient in wasingt''}  \\
\textbf{Proposed:}  
\textit{``i can say this is tomes gibs ge my one and only child and when he finished hiyschoul we had always pland to send him to prinstomn but his father had been called back into the servis as a reserve offiser and was stationd in washington''}  

The proposed model reduces errors substantially, improving WER from 40.43\% to 25.53\% (36.84\% relative improvement). Beyond the numerical score, the transcription from the proposed model is much more legible and semantically faithful to the reference, whereas the baseline contains multiple word merges and unintelligible segments. At 10 minutes of training data, the baseline model produced mostly gibberish, making detailed analysis impractical, whereas the proposed model yielded partially intelligible output.
\vspace{-10pt}
\section{Discussion and Conclusion}
\label{sec:conclusion}

Our experiments demonstrate that incorporating articulatory supervision through a multi-task framework yields consistent improvements in ASR performance across both \textit{base} and \textit{large} Wav2Vec2.0 configurations. The gains are particularly pronounced in the low-resource regimes, where relative WER reductions exceed 25\%, but remain measurable even with larger training sets and larger encoder models. Qualitative analysis further shows that the proposed approach improves transcription legibility, reducing merged words and unintelligible segments compared to the baseline. 

An additional observation is that the proposed \textit{base} model, when augmented with articulatory supervision, closely matches the performance of the much larger \textit{large} baseline model in the medium-resource settings (1 and 10 hours). In effect, adding articulatory information as both an auxiliary task and an input stream narrows the capacity gap, providing performance comparable to a model three times its size with only minimal added complexity. This highlights not only the accuracy benefits of our approach but also its efficiency. 


Looking ahead, we plan to explore integrating articulatory supervision into larger-scale models such as Whisper. This is challenging, since Whisper's decoder functions as an implicit language model, making it more difficult to late-fuse articulatory features with acoustic embeddings. Nevertheless, we believe that articulatory supervision could provide complementary benefits in this setting, possibly reducing hallucinations. Our ongoing work will investigate architectures that allow articulation to interact more naturally with Whisper.

\bibliographystyle{IEEEbib}
\bibliography{strings,refs}

\begin{thebibliography}{10}

\bibitem{benway2025perceptual}
Nina~R Benway, Saba Tabatabaee, Dongliang Wang, Benjamin Munson, Jonathan~L Preston, and Carol Espy-Wilson,
\newblock ``Perceptual ratings predict speech inversion articulatory kinematics in childhood speech sound disorders,''
\newblock {\em arXiv preprint arXiv:2507.01888}, 2025.

\bibitem{benway2025subtyping}
Nina~R Benway, Saba Tabatabaee, Benjamin Munson, Jonathan Preston, and Carol Espy-Wilson,
\newblock ``Subtyping speech errors in childhood speech sound disorders with acoustic-to-articulatory speech inversion,''
\newblock in {\em Proc. Interspeech 2025}, 2025, pp. 2800--2804.

\bibitem{10.1145/3462244.3479967}
Yashish~M. Siriwardena, Carol Espy-Wilson, Chris Kitchen, and Deanna~L. Kelly,
\newblock ``Multimodal approach for assessing neuromotor coordination in schizophrenia using convolutional neural networks,''
\newblock in {\em Proceedings of the 2021 International Conference on Multimodal Interaction}, New York, NY, USA, 2021, ICMI '21, p. 768–772, Association for Computing Machinery.

\bibitem{siriwardena2021inverted}
Yashish~M Siriwardena, Chris Kitchen, Deanna~L Kelly, and Carol Espy-Wilson,
\newblock ``Inverted vocal tract variables and facial action units to quantify neuromotor coordination in schizophrenia,''
\newblock in {\em Proc. 12th International Seminar on Speech Production (ISSP 2020)}, 2021, pp. 174--177.

\bibitem{11011243}
Gowtham Premananth and Carol Espy-Wilson,
\newblock ``Speech-based estimation of schizophrenia severity using feature fusion,''
\newblock in {\em 2025 IEEE International Conference on Acoustics, Speech, and Signal Processing Workshops (ICASSPW)}, 2025, pp. 1--5.

\bibitem{premananth2025multimodal}
Gowtham Premananth, Philip Resnik, Sonia Bansal, Deanna~L Kelly, and Carol Espy-Wilson,
\newblock ``Multimodal biomarkers for schizophrenia: Towards individual symptom severity estimation,''
\newblock {\em arXiv preprint arXiv:2505.16044}, 2025.

\bibitem{speech_synthesis_1}
Zhen-Hua Ling, Korin Richmond, and Junichi Yamagishi,
\newblock ``Articulatory control of hmm-based parametric speech synthesis using feature-space-switched multiple regression,''
\newblock {\em IEEE Transactions on Audio, Speech, and Language Processing}, vol. 21, no. 1, pp. 207--219, 2013.

\bibitem{speech_synthesis_2}
Korin Richmond and Simon King,
\newblock ``Smooth talking: Articulatory join costs for unit selection,''
\newblock in {\em 2016 IEEE International Conference on Acoustics, Speech and Signal Processing (ICASSP)}, 2016, pp. 5150--5154.

\bibitem{mitra2010articulatory}
Vikramjit Mitra, Hosung Nam, Carol~Y Espy-Wilson, Elliot Saltzman, and Louis Goldstein,
\newblock ``Articulatory information for noise robust speech recognition,''
\newblock {\em IEEE Transactions on Audio, Speech, and Language Processing}, vol. 19, no. 7, pp. 1913--1924, 2010.

\bibitem{mitra2017hybrid}
Vikramjit Mitra, Ganesh Sivaraman, Hosung Nam, Carol Espy-Wilson, Elliot Saltzman, and Mark Tiede,
\newblock ``Hybrid convolutional neural networks for articulatory and acoustic information based speech recognition,''
\newblock {\em Speech Communication}, vol. 89, pp. 103--112, 2017.

\bibitem{radford2023robust}
Alec Radford, Jong~Wook Kim, Tao Xu, Greg Brockman, Christine McLeavey, and Ilya Sutskever,
\newblock ``Robust speech recognition via large-scale weak supervision,''
\newblock in {\em International conference on machine learning}. PMLR, 2023, pp. 28492--28518.

\bibitem{baevski2020wav2vec}
Alexei Baevski, Yuhao Zhou, Abdelrahman Mohamed, and Michael Auli,
\newblock ``wav2vec 2.0: A framework for self-supervised learning of speech representations,''
\newblock {\em Advances in neural information processing systems}, vol. 33, pp. 12449--12460, 2020.

\bibitem{Westbury1994a}
John~R Westbury,
\newblock ``{Speech Production Database User ' S Handbook},''
\newblock {\em IEEE Personal Communications - IEEE Pers. Commun.}, vol. 0, no. June, 1994.

\bibitem{Schonle1987}
Paul~W. Sch{\"{o}}nle, Klaus Gr{\"{a}}be, Peter Wenig, J{\"{o}}rg H{\"{o}}hne, J{\"{o}}rg Schrader, and Bastian Conrad,
\newblock ``{Electromagnetic articulography: Use of alternating magnetic fields for tracking movements of multiple points inside and outside the vocal tract},''
\newblock {\em Brain and Language}, vol. 31, no. 1, pp. 26--35, may 1987.

\bibitem{parikh2022acoustic}
Rahil Parikh, Nadee Seneviratne, Ganesh Sivaraman, Shihab Shamma, and Carol Espy-Wilson,
\newblock ``Acoustic to articulatory speech inversion using multi-resolution spectro-temporal representations of speech signals,''
\newblock {\em arXiv preprint arXiv:2203.05780}, 2022.

\bibitem{siriwardena2022acoustic}
Yashish~M Siriwardena, Ganesh Sivaraman, and Carol Espy-Wilson,
\newblock ``Acoustic-to-articulatory speech inversion with multi-task learning,''
\newblock {\em arXiv preprint arXiv:2205.13755}, 2022.

\bibitem{7178964}
Vassil Panayotov, Guoguo Chen, Daniel Povey, and Sanjeev Khudanpur,
\newblock ``Librispeech: An asr corpus based on public domain audio books,''
\newblock in {\em 2015 IEEE International Conference on Acoustics, Speech and Signal Processing (ICASSP)}, 2015, pp. 5206--5210.

\bibitem{kendall2018multi}
Alex Kendall, Yarin Gal, and Roberto Cipolla,
\newblock ``Multi-task learning using uncertainty to weigh losses for scene geometry and semantics,''
\newblock in {\em Proceedings of the IEEE conference on computer vision and pattern recognition}, 2018, pp. 7482--7491.

\bibitem{tabatabaee2025enhancing}
Saba Tabatabaee, Suzanne Boyce, Liran Oren, Mark Tiede, and Carol Espy-Wilson,
\newblock ``Enhancing acoustic-to-articulatory speech inversion by incorporating nasality,''
\newblock {\em arXiv preprint arXiv:2506.09231}, 2025.

\end{thebibliography}

\end{document}